\documentclass[%
 reprint,
superscriptaddress,
 amsmath,amssymb,
 aps,
]{revtex4-1}

\usepackage{graphicx}
\usepackage{dcolumn}
\usepackage{bm}
\usepackage{hyperref}

\begin{document}

\preprint{APS/123-QED}

\title{Microrheology with Optical Tweezers:\\Measuring the solutions' relative viscosity \emph{at a glance}}

\author{Francesco Del Giudice}
\affiliation{Center for Advanced Biomaterials for Health Care @CRIB, IIT, P.le Tecchio 80, 80125 Naples, Italy.}

\author{Andrew Glidle}
\affiliation{Division of Biomedical Engineering, School of Engineering, University of Glasgow, Glasgow G12 8LT, UK}

\author{Francesco Greco}
\affiliation{Istituto di Ricerche sulla Combustione, IRC-CNR, P.le Tecchio 80, 80125 Naples, Italy.}

\author{Paolo~Antonio~Netti}
\affiliation{Center for Advanced Biomaterials for Health Care @CRIB, IIT, P.le Tecchio 80, 80125 Naples, Italy.}

\author{Pier~Luca~Maffettone}
\affiliation{Dipartimento di Ingegneria Chimica, dei Materiali e della Produzione Industriale, Universit$\acute{a}$ di Napoli Federico II, P.le Tecchio 80, 80125 Naples, Italy.}

\author{Jonathan M. Cooper}
\affiliation{Division of Biomedical Engineering, School of Engineering, University of Glasgow, Glasgow G12 8LT, UK}

\author{Manlio Tassieri}
\email{Corresponding author: Manlio.Tassieri@glasgow.ac.uk}
 \affiliation{Division of Biomedical Engineering, School of Engineering, University of Glasgow, Glasgow G12 8LT, UK}

\date{\today}

\begin{abstract}
We present a straightforward method for measuring the fluids' relative viscosity \textit{via} a simple graphical analysis of the normalised position autocorrelation function of an optically trapped bead, without the need of embarking on laborious calculations.~The advantages of the proposed microrheology method become evident, for instance, when it is adopted for measuring the molecular weight of rare or precious materials by means of their intrinsic viscosity.~The proposed method has been validated by direct comparison with conventional bulk rheology methods.
\end{abstract}

\pacs{Valid PACS appear here}
\maketitle



The pioneering studies of Albert~Einstein~\cite{ISI:000201930400011}, at the beginning of the twentieth century, introduced what would become one of the most important parameters in the field of rheology: the solution \textit{relative viscosity} ($\eta_r$), defined as the ratio of the solution viscosity ($\eta$) to that of the solvent ($\eta_s$). Einstein derived this expression for a suspension of hard spheres at low volume fractions (i.e. $\phi \lesssim 1\%$): $\eta_r=1+2.5\phi$.
The latter was the spark that led to a myriad of studies (e.g.~\cite{ISI:A1977DZ60100006, ISI:A1959XG59900012, ISI:000274932000014, ISI:000296985400002}) seeking to find the \emph{yet undefined} laws governing the rheology of highly concentrated (i.e. for $\phi \gg 1\%$) suspensions.

Later, with the advent of polymer physics (followed by the birth of the field of rheology in the~$1929$), scientists established that for very dilute polymer solutions, the viscosity increases above the solvent viscosity linearly with the polymer concentration, $c$, and that the effective `\emph{virial expansion}' for viscosity is: $\eta=\eta_s(1+\left[\eta\right]c+k_H\left[\eta\right]^2c^2+\cdots)$, where $\left[\eta\right]$ is the intrinsic viscosity and $k_H$ is the Huggins coefficient~\cite{Ferry:1980jo}. The intrinsic viscosity can be seen as either the initial slope of the relative viscosity or as the linear extrapolation to zero concentration of the reduced viscosity $\eta_{red}=(\eta_r-1)/c$ when these are plotted against mass concentration.

The ability to determine the polymers' intrinsic viscosity from rheological measurements became of great interest to a broad scientific community when it was found that $\left[\eta\right]$ is simply related to the polymer molecular weight ($M$) by means of the Mark--Houwink equation: $\left[\eta\right]=KM^\alpha$, where $K$ and $\alpha$ are two constants that are tabulated for nearly all linear polymers in various solvents~\cite{Rubinstein:2003fb}.

Conventionally, there are two popular methods for measuring $\eta_r$: the first, which is also the most accurate, is performed by using an Ubbelohde viscometer (named after its inventor Leo Ubbelohde), which requires the measurement of the liquids' efflux time through a thin capillary of known geometries; the ratio between the measured times of a pair of fluids is simply proportional to their $\eta_r$~\cite{Viswanath:2007aa}. The second method involves the measurement of the liquids' steady speed of deformation (i.e.~the shear rate $\dot{\gamma}$) occurring as consequence of a known applied constant stress; the ratio between the measured shear rates provides a measure of the relative viscosity of two fluids, if the stress is kept the same in both the measurements.

Despite their simplicity, both of the above methods require millilitres of sample volume (i.e.~of the order of tens of $ml$), resulting in being unsuitable for rare or precious materials such as those involved in biological studies~\cite{ISI:000329157000012, ISI:000328623600017}.~This emphasises the importance of the \emph{novel} microrheology methods with optical tweezers (OT) for measuring $\eta_r$.~Indeed, like the other microrheology techniques~\cite{ISI:000228157400004}, it only requires few micro-litres of sample volume per measurement, \emph{plus} it provides a straightforward and accurate procedure for measuring the solutions' relative viscosity -- and therefore the materials' molecular weight \textit{via} their intrinsic viscosity. This can be achieved by means of a simple \emph{visual} analysis of the normalised position autocorrelation function of an optically trapped bead, with the added advantage of avoiding laborious calculations~\cite{ISI:A1995QG47200053, ISI:000088854900006, ISI:000176552300053, Evans:2009dw, T12}.

Optical tweezers have thus proved to be both a valuable and versatile tool for microrheology purposes~\cite{ISI:000275053800036, ISI:000291926500023, T12, ISI:000322416700024}. In particular, it has been shown~\cite{T12} that the statistical mechanics analysis of the trajectory of an optically trapped bead is able to uncover the fluids' linear viscoelastic properties (over a \emph{wide} range of frequencies) \textit{via} the Fourier transform of either of the following two time-averaged quantities:~the normalised mean-square displacement (NMSD) $\Pi(\tau)=\langle \left[\vec{r}(t+\tau)-\vec{r}(t)\right]^{2}\rangle/2\langle r^2 \rangle$ \textit{or} the normalised position autocorrelation function (NPAF) $A(\tau)=\langle \vec{r}(t)\vec{r}(t+\tau)\rangle/\langle r^2 \rangle$; where $\vec{r}$ is the particle position from the trap centre $\vec{r}_0 \equiv \vec{0}$, $\tau$ is the lag-time (or time interval), $\langle r^2 \rangle$ is the time-independent variance of the bead position. Note that, the NMSD and the NPAF are simply related to each other as $\Pi(\tau)+A(\tau)=1$.

The fluid's linear viscoelastic properties can be represented by its complex shear modulus $G^*(\omega)=G'(\omega)+iG''(\omega)$, which is a complex number whose real and imaginary parts, named as the storage and the loss modulus, provide information on the elastic and the viscous nature of the fluid, respectively. Note that $G^*(\omega)$ is time invariant~\cite{Ferry:1980jo}.
Tassieri~\textit{et al.} have showed that there is a straightforward relationship between the above quantities, i.e. Eq. (7) of Ref.~\cite{T12}, here expressed as:
\begin{equation}
G^*(\omega)\frac{6\pi a}{\kappa}=\frac{\hat{A}(\omega)}{\hat{\Pi}(\omega)}
\label{G*}
\end{equation}
where $\hat{\Pi}(\omega)$ and $\hat{A}(\omega)$ are the Fourier transforms of $\Pi(\tau)$ and $A(\tau)$, respectively, $a$ is the bead radius and $\kappa$ is the OT trap stiffness, which can be easily determined by appealing to the Principle of Equipartition of Energy: $k_{B}T=\kappa \langle x^2 \rangle$ written in one dimension.~The evaluation of the Fourier transforms in Eq.~(\ref{G*}), given only a finite set of data points over a finite time domain, is non-trivial since interpolation and extrapolation from these data can yield artifacts that lie within the bandwidth of interest~\cite{Evans:2009dw, T12}. Nevertheless, although the above issue has been solved~\cite{T12}, we wish to demonstrate that an important rheological parameter such as the solution relative viscosity can be determined \textit{via} a simple visual analysis of $A(\tau)$; without the need of performing a Fourier transform.

In order to validate the method, we have determined the molecular weight of two known polyacrylamides (PAMs), having molecular weights of~$1.5~kDa$ and $1.145~MDa$, both from Polysciences Inc., using their intrinsic viscosity. The latter has been extrapolated from relative viscosity measurements performed with OT on water-based solutions of PAM at different concentrations, as described hereafter.

\begin{figure}[t]
\begin{center}
\scalebox{0.46}{\includegraphics{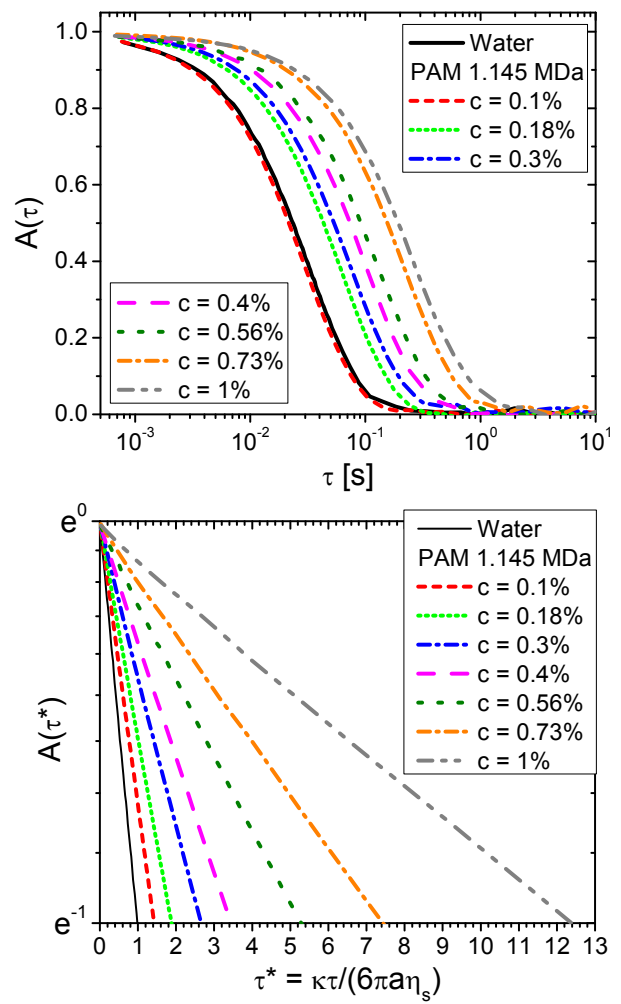}}
\caption{(Top) Linear-Log plot of the normalised position autocorrelation function $A(\tau)$ \textit{vs.} lag-time of a $5\mu m$ diameter bead suspended in water and water-based solutions of polyacrylamide ($M=1.145~MDa$) at concentrations ranging from $0.1\%~w/w$ to $1\%~w/w$. (Bottom) Ln-linear plot of the same data as shown above but drawn \textit{vs.}~a dimensionless lag-time $\tau^*=\tau\kappa / (6\pi a \eta_s)$, where $\eta_s$ is the solvent viscosity, here water ($\eta_s=0.896 mPa\cdot s$). The scale of the ordinate axis has been limited to the region of interest, i.e. $A(\tau^*)\in\left[e^{-1},1\right]$. Notably, for the case of pure water the abscissa of the intercept of $e^{-1}$ with $A(\tau^*)$ is $1$.}
\label{Fg_1}
\end{center}
\end{figure}

When an optically trapped particle is suspended in a Newtonian fluid (i.e.~a fluid with constant viscosity $\eta$), $A(\tau)$ assumes the form of a single exponential decay~\cite{T12}:
\begin{equation}
A(\tau)=e^{-\lambda \tau}
\label{A1}
\end{equation}
where $\lambda=\kappa / (6\pi a \eta)$ is the characteristic relaxation rate of the compound system (i.e. OT, bead and fluid). From Eq.~(\ref{A1}), it is a matter of a simple change of variables to show that:
\begin{equation}
A(\tau^*)=e^{-\tau^*/\eta_r}
\label{Ar}
\end{equation}
where $\tau^*=\tau\kappa / (6\pi a \eta_s)$ is a dimensionless lag-time and $\eta_s$ is the solvent viscosity, assumed to be Newtonian (here this condition is fulfilled because the solvent is water).~It follows that, by drawing a horizontal line starting from the ordinate $e^{-1}$, the abscissa of its intercept with $A(\tau^*)$ provides a \emph{reading} of the solution relative viscosity $\eta_r$, as shown in Figure~\ref{Fg_1} for water and water-based solutions of polyacrylamide ($M=1.145~MDa$) at concentrations ranging from $0.1\%~w/w$ to $1\%~w/w$.~Notably, in the case of pure water, the abscissa of the intercept of $e^{-1}$ with $A(\tau^*)$ is $1$.

It is important to highlight that, in general, polymer solutions are non-Newtonian, especially at relatively high concentrations and Eq.~(\ref{A1}) may not be valid, at least not for all concentrations.
However, in dilute conditions (i.e., at relatively low polymer concentrations) most of solutions tend to show a Newtonian behaviour, especially towards vanishingly small values of concentration, which \emph{coincidentally} are the same conditions required for measuring $\left[\eta\right]$. Hence the applicability of Equations~(\ref{A1})~and~(\ref{Ar}) for measuring $\eta_r$.

\begin{figure}[b]
\begin{center}
\scalebox{0.45}{\includegraphics{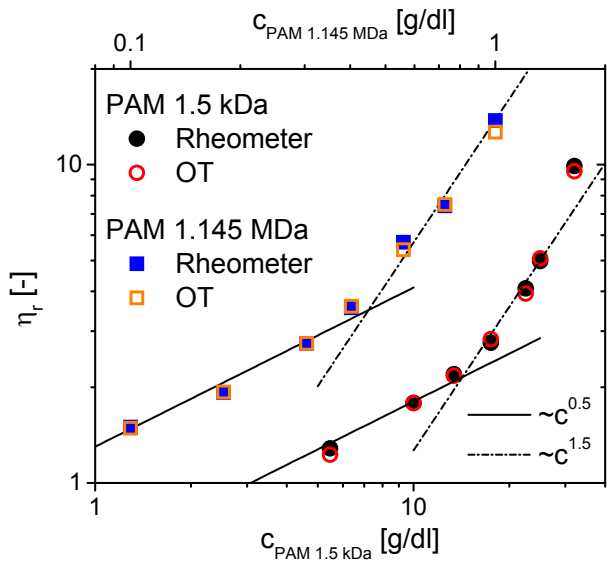}}
\caption{The relative viscosity \textit{vs.}~concentration of two water-based solutions of polyacrylamides having molecular weights of $1.5~kDa$ (bottom axis) and $1.145~MDa$ (top axis). The filled and the open symbols refer to bulk- and micro-rheology measurements of $\eta_r$, respectively. The lines are guides for the gradients.}
\label{Fg_2}
\end{center}
\end{figure}

Figure~\ref{Fg_2} shows a comparison between the relative viscosities of two sets of PAMs solutions measured with both OT, as described above, and with a conventional stress-controlled rheometer (Anton-Paar MCR-302) equipped with a cone and plate geometry of $50~mm$ diameter and $1^o$ angle.~The agreement between micro- and bulk-rheology is shown.~Moreover, for both the PAMs, the solutions' viscosities obey the theoretical predictions of the concentration scaling-laws for linear polyelectrolytes~\cite{Rubinstein:2003fb}: i.e., $\eta_r \propto c^{0.5}$ for semi-dilute regime and $\eta_r \propto c^{1.5}$ for entangled regime; with the identification of the entanglement concentration $c_e$ at the transition of the two regimes, for both the PAMs.

\begin{figure}[t]
\begin{center}
\scalebox{0.43}{\includegraphics{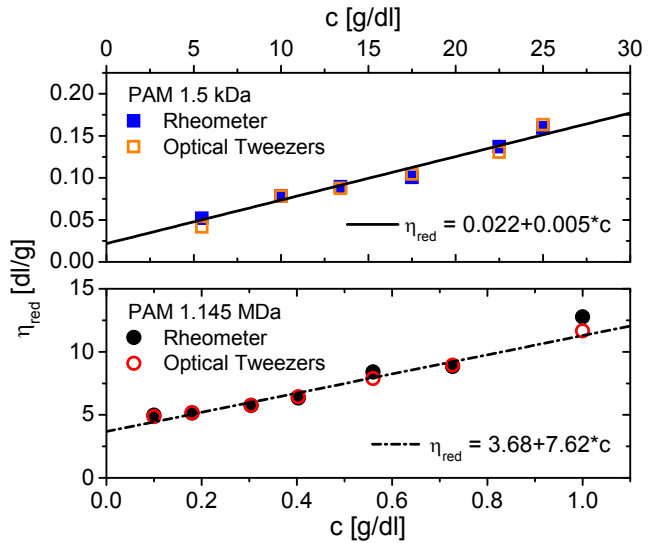}}
\caption{The reduced viscosity \textit{vs.}~concentration derived from the data shown in Fig.~(\ref{Fg_2}). The filled and the open symbols refer to bulk- and micro-rheology measurements of $\eta_{red}$, respectively. The lines are linear fits of OT data; the extrapolations to zero concentration of both the fits provide a \emph{reading} of $\left[\eta\right]$ for both the PAMs.}
\label{Fg_3}
\end{center}
\end{figure}

Once the solutions' relative viscosities are known, it is a simple step to reorganise the data in terms of their reduced viscosities, as shown in Figure~\ref{Fg_3}. The linear extrapolation to zero concentration of $\eta_{red}$ provides a \emph{reading} of the PAMs' intrinsic viscosities. These, as introduced earlier, are simply related to the polymers molecular weight by means of the Mark--Houwink equation, which for water-based solutions of polyacrylamides writes~\cite{ISI:A1980KE43700020}:
\begin{equation}
\left[\eta\right]=6.31\times 10^{-3}M^{0.80}
\label{EIL}
\end{equation}
\begin{equation}
\left[\eta\right]=4.90\times 10^{-3}M^{0.80}
\label{EIH}
\end{equation}
where the above two equations are supposed to be valid for PAMs with low and high molecular weights, respectively.
In Table~\ref{tab:M} we report the results obtained by substituting the values of $\left[\eta\right]$ derived from Figure~\ref{Fg_3} in both Equations~(\ref{EIL}) and~(\ref{EIH}) and compare them with the nominal values of the molecular weights provided by the supplier; the agreement is very good, especially when considered the respective range of validity of the two equations.

In conclusion, we have introduced and validated a simple experimental method for measuring the solutions' relative viscosity with optical tweezers, by means of a visual analysis of the particle normalised position autocorrelation function. The advantages of the proposed method rely not only on its simplicity, but also on its microrheology nature (i.e. it requires micro-litres sample volume), which makes it of great interest to all those studies where rare and precious materials are involved (such as biological studies).

We thank Mike Evans and Miles Padgett for helpful conversations. MT acknowledge support via personal research fellowships from the Royal Academy of Engineering/EPSRC.

\begin{table}[b]
	\medskip
	\begin{tabular}{| >{\centering\arraybackslash}p{1.5cm} | >{\centering\arraybackslash}p{1.8cm} | >{\centering\arraybackslash}p{1.5cm} | >{\centering\arraybackslash}p{2cm} |}
	\hline
	$\left[\eta\right]$ & Equation & $M$ & $M$ \\
	($dl/g$)  & -- & ($Da$) & ($Da$) \\
	 \hline\hline
		-- & \textit{Nominal} & $\textit{1,500}$ & $\textit{1,145,000}$ \\
		$0.022$ & Eq.~(\ref{EIL}) & $\textbf{1,506}$ & $906,302$ \\
		$3.68$ & Eq.~(\ref{EIH}) & $2,066$ & $\textbf{1,240,000}$ \\
		\hline
	\end{tabular}
	\caption{\label{tab:M}Comparison between nominal and measured molecular weights of two commercially available polyacrylamides. Equations~(\ref{EIL}) and~(\ref{EIH}) have been taken from Ref.~\cite{ISI:A1980KE43700020}.}
\end{table}

\clearpage

\bibliographystyle{unsrt}

\bibliography{ReViOT}

\end{document}